\documentclass[runningheads]{llncs}
\usepackage{graphicx}
\usepackage{xcolor}
\usepackage{subfigure}
\usepackage{subfig}
\begin{document}
\title{How to Achieve High Classification Accuracy with Just a Few Labels: A Semi-supervised Approach Using Sampled Packets
}
\titlerunning{How to Achieve High Classification Accuracy with Just a Few Labels}
%
\author{Shahbaz Rezaei \and
Xin Liu}
\authorrunning{Industrial Conference on Data Mining (ICDM) 2019, New York, USA}
%
\institute{University of California, Davis, CA 95616, USA \\
\email{\{srezaei,xinliu\}@ucdavis.edu}}
\maketitle              
\begin{abstract}
Network traffic classification, which has numerous applications from security to billing and network provisioning, has become a cornerstone of today's computer networks. Previous studies have developed traffic classification techniques using classical machine learning algorithms and deep learning methods when large quantities of labeled data are available.
However, capturing large labeled datasets is a cumbersome and time-consuming process. In this paper, we propose a semi-supervised approach that obviates the need for large labeled datasets. We first pre-train a model on a large unlabeled dataset where the input is the time series features of a few sampled packets. Then the learned weights are transferred to a new model that is re-trained on a small labeled dataset. We show that our semi-supervised approach achieves almost the same accuracy as a fully-supervised method with a large labeled dataset, though we use only 20 samples per class. Duting inferene, based on a dataset generated from the more challenging QUIC protocol, our approach yields   98\% accuracy. To show its efficacy, we also test our approach on two public datasets. Moreover, we study three different sampling techniques and demonstrate that sampling packets from an arbitrary portion of a flow is sufficient for classification.

\keywords{Transfer Learning \and Semi-supervised Learning \and Network Traffic Classification \and Packet Sampling \and QUIC Protocol Classification}
\end{abstract}

\section{Introduction}\label{sec:intro}
Network traffic classification is one of the key components of network management and administration systems.  It has been used for Quality of Service (QoS) provisioning, pricing, anomaly detection, malware detection, etc. Depending on the usage scenario, traffic may be classified based on protocols (e.g. UDP, TCP or FTP), traffic-types (e.g. voice, video or downloading), application (e.g. Youtube, Facebook or WeChat), user-actions, operating system, etc \cite{rezaei2018deep}. Due to the inherent differences of traffic in these applications, a model, approach or dataset that is used for one application cannot be used for another application. As a result, due to the lack of general approach, designing an accurate traffic classifier for an application of interest a has been time-consuming and cumbersome task.

Classical machine learning approaches have been extensively used for more than a decade and showed good  accuracy. However, the emergence of new applications and encryption protocols has dramatically increased the complexity of the traffic classification problem.
Recently, deep learning algorithms  have been developed  for traffic classification. Deep learning approaches are capable of automatic feature selection and capturing highly complex patterns, and thus demonstrated high classification accuracy in comparison to  other methods. 

However, deep learning  requires a large amount of labeled data during training. Capturing and labeling a large dataset is a non-trivial and cumbersome process. First, current Internet traffic is mostly encrypted that makes DPI (Deep Packet Inspection) based labeling impossible. Hence, most labeling procedures  capture each traffic class in isolation. However, this is only possible at the edge of a network or in an isolated environment. Unlike labeled dataset, unlabeled data is abundant and readily available in the Internet. Therefore, it motivates us to study how to use easily-obtainable unlabeled datasets to dramatically reduce the size of labeled dataset needed for accurate traffic classification.

Furthermore, to make our approach practical, in particular for high speed links or data centers, we propose to use sampled data packets instead of an entire flow. 
Using sampled packets also reduces memory and computation complexity needed for entire time series features \cite{rezaei2018deep}. In summary, in this paper, we make the following contributions:
\begin{enumerate}
  \item We propose a semi-supervised approach that utilizes large quantities of unlabeled data and just a few labeled samples.  Specifically, we first train a model on a large unlabeled dataset and then re-train the model with a few labeled data on the target classification problem.
  \item We study three different sampling methods on the encrypted traffic classification problem: Fixed step sampling, random sampling, and incremental sampling. We show that good sampling methods can almost achieve the upper-bound accuracy in certain datasets.
  \item We evaluate the proposed approach using captured QUIC traffic and show that our semi-supervised approach using sampled packets can accurately classify QUIC traffic that has fewer unencrypted fields during handshake.
  \item We test our approach on different public datasets. Surprisingly, we show that we can train a model using a completely separate unlabeled dataset and then retrain the model with a small number of labels in the target dataset and still achieve good  accuracy.
\end{enumerate}

\section{Related Work}
In pre-deep learning era, traditional machine learning algorithms had been commonly used for traffic classification \cite{velan2015survey}. However, due to their simplicity, manual feature extraction, and inability to capture complex patterns, their accuracy has declined \cite{rezaei2018deep}. Recently,
several studies develop deep learning models, such as Convolutional Neural Network (CNN) and Long Short-Term Memory (LSTM), for network traffic classification in a fully-supervised fashion. In \cite{zhou2017method}, authors train six different CNN models based on LeNet-5 model on a public dataset with 12 classes. They convert 249 statistical characteristics into a 2-d 16$\times$16 image and report high accuracy. In \cite{lopez2017network}, authors use CNN, LSTM and various combinations on a private dataset captured at RedIRIS, a Spanish academic and research backbone network. They use time series features of the first 20 packets, including source port, destination port, payload size, window size, etc.

In \cite{lotfollahi2017deep}, a framework comprising a CNN and Stacked Auto-Encoder (SAE) is trained on a dataset containing 12 VPN and non-VPN traffic classes. They use raw header and payload bytes as input. In \cite{chen2017seq2img}, Reproducing Kernel Hilbert Space (RKHS) is used to convert the time series features of a flow to an image. Then, produced images are used as input to a CNN model. The only study that investigate QUIC protocol is \cite{tong2018novel}. They capture five Google services: Google Hangout Chat, Google Hangout Voice
Call, YouTube, File transfer, and Google play music. They use CNN model and report high accuracy. They capture 150GB of data and train the model in a fully-supervised manner.

In \cite{rezaei2018deep}, a general framework, covering all previous deep learning-based traffic classifiers, is introduced that can deal with any typical network traffic classification task. The paper provide a seven-step training process, including data capturing, data pre-processing, model selection and evaluation, etc. The framework also requires large enough dataset for training.  In summary, in comparison, all work discussed above assumes large quantities of labeled data. Furthermore, packet sampling is not considered in their approaches.



\section{Problem Statement}

As discussed earlier, deep learning models have been adopted for (encrypted) traffic classification. Because deep learning approaches are capable of automatic feature selection and capturing highly complex patterns, they demonstrate high classification accuracy. However, a critical challenge is that deep models require large amounts of labeled data during training. Capturing and labeling a dataset is a non-trivial and cumbersome process.

First, because encryption mechanisms are heavily used in today's Internet, labeling a dataset captured in an operational network is almost impossible unless an accurate classifier is already available. As a result, labeling process is usually done by assuring that only one target class is available during capturing by turning off all other applications on the device used to generate traffic. This is typically done in an isolated environment or at the edge of the network. Traffic distribution in such an environment may differ from an operational network, especially at the core of the network. In addition, to capture large amounts of labeled data, previous studies often runs scripts to automatically perform certain actions that can be captured and labeled without manual labor. However, we will show that network traffic generated by scripts may have a different distribution from that of human-generated traffic, causing poor performance on real traffic.

In contrast, unlabeled data is abundant and readily available in an operational network. Capturing a large unlabeled dataset is an easy task. There are also many large and publicly available datasets. Hence, our objective is to use easily-obtainable unlabeled datasets to significantly reduce the size of labeled dataset needed for training an accurate traffic classifier. We only have a few labeled samples for each class that we are interested in, called target classes. We call this dataset $D_l$. At the same time, we assume we have a large unlabeled dataset $D_u$. This unlabeled dataset $D_u$ may contain numerous flows from various traffic classes, even from classes that we are not interested in, i.e. they do not exist in $D_l$. The goal is to leverage the two datasets to train a traffic classification model for a target task while heavily exploiting $D_u$ dataset for training.

\section{Methodology}

\subsection{Key Components}

Our objective is to obtain an accurate traffic classifier with only a small number of labeled samples from each traffic class. To achieve this objective, we propose a semi-supervised learning approach. There are three key components in the proposed scheme. The first is the classification model trained through semi-supervised learning. Specifically, we  pre-train a model with $D_u$ and then transfer the model to a new architecture and re-trained the model with $D_l$. This is called semi-supervised learning. This approach  considerably reduces the number of labeled data needed for the second supervised learning part. Moreover, the pre-trained model can be reused for other network traffic classification tasks.

In order to use an unlabeled dataset, $D_u$, we pre-train a model, $F$, such that it does not need human labor for labeling\footnote{ We use pre-training and re-training to distinguish between the first and second step of our semi-supervised approach.}. An important step in the pre-training stage is to decide the target of the regression function. We choose a set of statistical features for this purpose, such as average packet length, average inter-arrival time, etc. Moreover, the input of the model is a set of packets sampled from the entire flow. For each packet, we only observe length, direction, and relative time. In other words, $F$ is pre-trained to estimate statistical features of the entire flow by taking a set of sampled packets as an input. The idea is based on the assumption that not all traffic patterns are valid and a model pre-trained on a large unlabeled dataset will lay on a manifold of valid patterns and hopefully can estimate statistical features.

Next, the pre-trained model, $F$, will be used as a part of another model, $G$. Then, G will be re-trained with a small labeled dataset. Since a part of the model has already observed many traffic patterns, $G$ needs considerably less human-labeled data. Note that $F$ might not necessarily be an accurate estimator of statistical features. But, it will be re-trained quickly to help the classification task since it has already seen numerous traffic patterns during the pre-training.

The second key component is the features, including input features and the pre-training targets.
As it is categorized in \cite{rezaei2018deep}, there are three input features heavily used for network traffic classification tasks: time series features (such as packet length and inter-arrival time), statistical features obtained from the entire flow (such as average packet length and average byte sent per second), and header/payload features (such as TCP window size field, TLS handshake data fields and data content). Header/payload data features have been used for classification of encrypted traffic such as TLS $1.2$ \cite{lotfollahi2017deep,shen2017classification}. These methods rely upon unencrypted data fields or message types exchanged during handshake phase of TLS $1.2$. However, state-of-the-art encryption protocols, such as QUIC and TLS $1.3$, aim to reduce the number of handshake messages to improve the speed of connection establishment. As a result, fewer unencrypted data fields and packets are exchanged that makes it harder for header/payload based approaches to achieve high classification accuracy. Furthermore, statistical features require the model to observe the entire flow before classification which is not efficient in practice. However, statistical features present useful information of flows, and thus we use statistic features as the target, not input, in the pre-training stage.

The third component of our approach is sampling. As discussed earlier, statistical features and header/payload features have their limitations. 
Therefore, we aim to use time series features. Specifically, we use time series features of only a part of a flow as an input to  predict the statistical features as a regression target. Additionally, instead of using only the first few packets, which is used in most studies based on time series features \cite{chen2017seq2img,conti2016analyzing,lopez2017network}, we sampled a fixed number of packets from a flow for the input. First, in practice, sampling is the only practical solution in some scenarios, such as in high bandwidth links or data centers. Moreover, sampling obviates the need to start capturing from the beginning of a flow or capturing the entire flow. Furthermore, storing the entire flow needs a large  amount of memory and a model trained on would be more complex. Additionally, it allows the model to capture patterns from different part of a flow, not just the beginning of a flow. The approaches that used the first few packets can only capture specific patterns taken place at the beginning of a connection. However, these patterns may not necessarily be distinguishable from one class to another. For instance, user-specific behaviors, such as changing the quality of Youtube video, renaming a file in a Google Drive, etc., are mostly performed at the middle of a flow, not within the first few packets. Finally, it allows us to sample a single flow several times which serve as a data augmentation method.


\begin{figure*}
	\centering
	\includegraphics[width=5in]{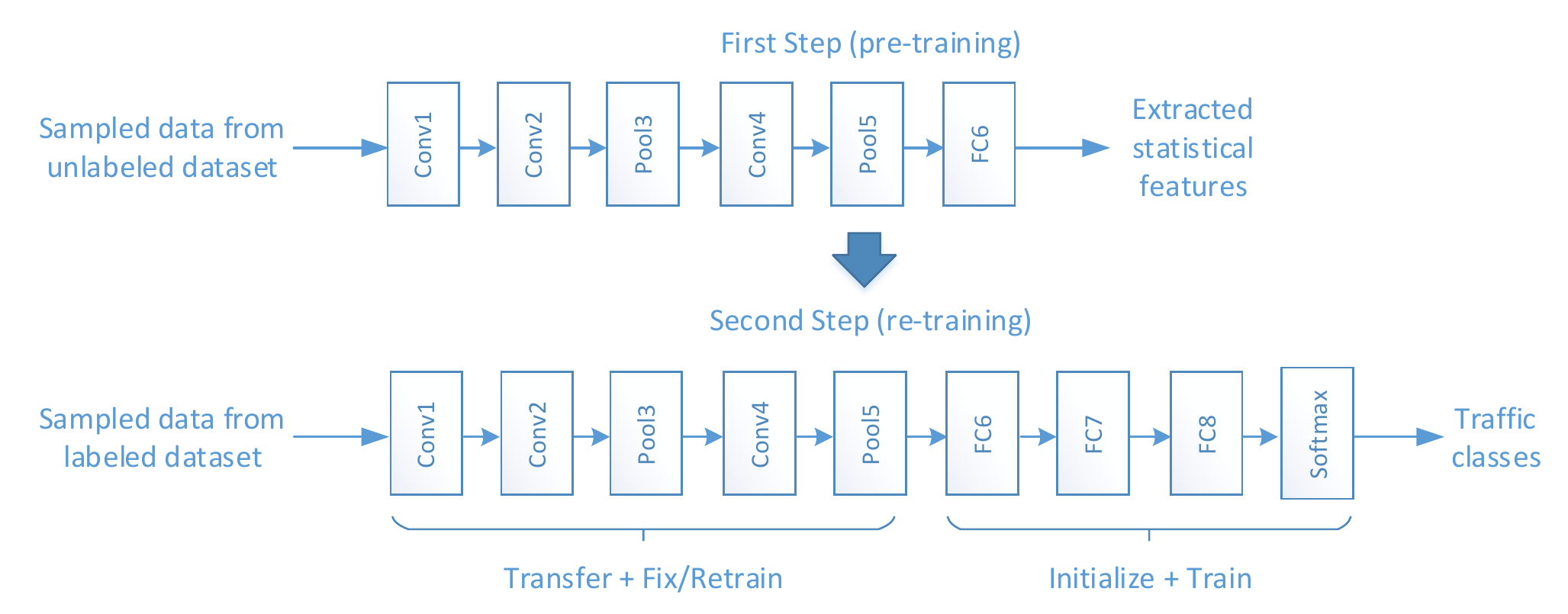}
	\caption{Semi-supervised steps and model architecture}
	\label{fig-model}
\end{figure*}

\subsection{Semi-supervised Learning}

\label{sec-model-arc}

We used Convolutional Neural Network (CNN) as a part of the model architecture because of its shift invariant feature. CCN uses same set of small filters to cover the entire receptive fields. Hence, it allows the model to be shift invariant, that is, a pattern can be captured by CNN even if it is shifted to another region of input. Recall that we used sampled packets as an input to the model and as a data augmentation technique we sampled multiple times from different part of the flow. Hence, same set of patterns may be observed in different part of the input. Thus, the shift invariant model is more suitable.

\begin{table*}[t]
	\centering
	\caption{Structure of the CNN model}
	\label{tbl-model-specification}
	\begin{tabular}{*{15}{c}}
		\hline
		- & Conv1 & Conv2 & Pool3 & Conv4 & Pool5 & FC6 & FC7 & FC8 \\
		\hline
		Number of filters/neurons & 32 & 32 & - & 64 & - & 256 & 128 & 128 \\
		\hline
		Kernel size & 5 & 5 & 3 & 3 & 3 & - & - & - \\
		\hline
	\end{tabular}
\end{table*}

As shown in Fig. \ref{fig-model}, a CNN-based model is first pre-trained with an unlabeled dataset. Then, the learned weights of the convolutional layers are transferred to a new model with more linear layers. 
Finally, the new model is re-trained on a small labeled dataset. The details of the model structure is presented in Table \ref{tbl-model-specification}. We use max pooling and Rectified Linear Unit (ReLU) activation function. Batch normalization is also used after convolutional and max pooling layers.

The input of the model is a 1-dimensional vector with 2 channels. The first channel contains inter-arrival time of sampled packets and the second channel contains the packet length and direction combined. To combine the packet length and direction, we multiply the direction (+1 or -1) with packet length. So, if packet length is positive, it shows packet length in forward direction, otherwise it shows the packet length in backward direction.
Moreover, we normalized the input data range in [-1,+1] by considering the maximum value of 1434 Bytes and 1 second for length and inter-arrival time. 

\subsection{Sampling Techniques}
In this paper, we used 3 different sampling methods to examine the effect of sampling on performance of the traffic classification task.
\begin{itemize}
	\item \textbf{Fixed step sampling:} In fixed step sampling, a fixed number, \textit{l}, is chosen and only packets that are \textit{l} packets away from are sampled.
	\item \textbf{Random sampling:} This technique simply samples each packet with probability \textit{p} $<$ 1. This is a common technique in operational networks with high bandwidth because it requires less memory and computational overhead.
	\item \textbf{Incremental sampling:} Incremental sampling has three parameters, (\textit{l, $\alpha$, $\beta$}). Similar to fixed step sampling, it samples packets that are \textit{l} packets away from, but it increases the value of \textit{l} by multiplying it by $\alpha$ after sampling each $\beta$ packets.
\end{itemize}

During data augmentation, we sampled a flow 100 times from the beginning of the flow when random sampling was used. However, it would have given us the same set of packets if we had started from the beginning of a flow multiple times when using fixed step or incremental sampling. Hence, we started sampling at different part of a flow 100 times, if the flow was long enough.

\subsection{Datasets}
As explained earlier, our semi-supervised approach needs an unlabeled dataset for the pre-training stage and a labeled dataset for the re-training stage. In this paper, we conducted experiments with three datasets:

\textbf{QUIC Dataset:}
This is a dataset captured in our lab at UC Davis and contains 5 Google services: Google Drive, Youtube, Google Docs, Google Search, and Google Music \cite{ourDataset}. We used several systems with various configurations, including Windows 7, 8, 10, Ubuntu 16.4, and 17 operating systems. We wrote several scripts using Selenium WebDriver \cite{selenium} and AutoIt \cite{autoit} tools to mimic human behavior when capturing data. This approach allowed us to capture a large dataset without significant human effort. Such approach has been used in many other studies \cite{tong2018novel,dubin2017know,conti2016analyzing}. Furthermore, we also captured a few samples of real human interactions to show how much the accuracy of a model trained on scripted samples will degrade when it is tested on real human samples. During preprocessing, we removed all non-QUIC traffic. Note that all flows in our dataset are labeled, but we did not use labels during the pre-training step. We used class labels of all flows to show the accuracy gap between a fully-supervised and semi-supervised approach.

\textbf{Unlabeled Waikato Dataset:}
WAND network research group at the university of Waikato published several unlabeled traces from 2009 to 2013. In this paper, we use Waikato VIII \cite{waikato} captured at the border of the University of Waikato. The entire dataset is unlabeled and it is not clear what traffic classes exist in the dataset. However, the dataset definitely do not contain QUIC traffics because it was captured before the emergence of any practical implementation of QUIC protocol. We use Waikato dataset to pre-train the CNN model.  The dataset is extremely large and due to the limited time and computational budget, we only used traces of the first month, around $4\%$ of the entire dataset.

\textbf{Ariel Dataset:}
Ariel dataset \cite{ariel} was captured in a research lab at Ariel university over a period of two months. The original paper \cite{muehlstein2017analyzing} used a fully-supervised method to classify three category of class labels: operating system, browser, and application. However, only the operating system and browser labels are available in the public dataset. In this paper, we only use a small portion of the dataset to re-trained a pre-trained model to test our methodology.

For all datasets, We ignore short flows because when short flows are sampled, there will not be enough packets to feed a classifier. In our evaluation, short flows are those with less than 100 packets before sampling.

\section{Evaluation}
\subsection{Implementation Detail}
We used python and implemented the CNN architecture using PyTorch. We used a server with Intel Xeon W-2155 and Nvidia Titan Xp GPU using Ubuntu 16.04. The CNN model has already been explained in section \ref{sec-model-arc}. 
During the pre-training, we trained the model with Adam optimizer and MSE loss function for 300 epochs. We used 24 statistical features as targets of regression. We used minimum, maximum, average, and standard deviation of packet length and inter-arrival time. For each one, we considered forward, backward and both flow directions that gave us a total of 24 features. During the supervised re-training, we used Adam optimizer with cross-entropy as a loss function\footnote{ Codes are available at \url{https://github.com/shrezaei/Semi-supervised-Learning-QUIC-}}. 

There are two category of performance measures to evaluate a classifier: macro-average and micro-average metrics \cite{van2013macro}. Whenever the accuracy of the entire model is shown, we used macro-averaging where the accuracy is averaged over all classes. For pre-class performance evaluation, we used micro-averaging metrics, including accuracy, precision, recall, and F1, similar to \cite{tong2018novel}.



\subsection{QUIC Dataset}
\label{sec-quic-performance}

We first pre-train our model on QUIC dataset consisting of 6439 flows without using class labels. Recall that because we sample each flow up to 100 times, total number of samples during the training is $544744$. We use $24$ statistical features calculated from flows as regression targets. Then, we transfer the weights to a new model and re-train with class labels. For this step, we capture 30 flows for each class and divide the training and test with different number of flows \footnote{ In the entire paper, we use flow to refer to an unsampled flow and sampled flow for the sampled case. In other words, a dataset with 30 flows per class contains up to 3000 sampled flows per class because each unsampled flow is sampled multiple times.}. We also perform cross-validation. Moreover, we train the same model without transferring the weight to show the performance gap.

To tune the hyper-parameters, we separate 30 files for each target class and conduct greedy search. We train the model  with only 20 labeled flows and validate with other 10 flows. This small number of training data may not yield optimal parameters, but it is consistent with the assumption that a limited amount of labeled data is available for the supervised training we conduct. We find that the model shown in Fig. \ref{fig-model} is accurate enough and a deeper model does not yield higher accuracy. We also find that re-training or fixing the convolutional part of the transferred model during re-training does not significantly change the accuracy. Note that we use these same hyper-parameters for other experiments without re-tuning them. Hence, these sets of hyper-parameters seem to be acceptable across different datasets, which makes  our proposed approach more practical. The sampling parameters we use for fixed, incremental and random sampling are $22$, $(22, 1.6, 10)$, and $1/22$, respectively. We also sample only 45 packets from the entire flow.



\begin{figure}
\subfigure[]{\includegraphics[width=2.6in]{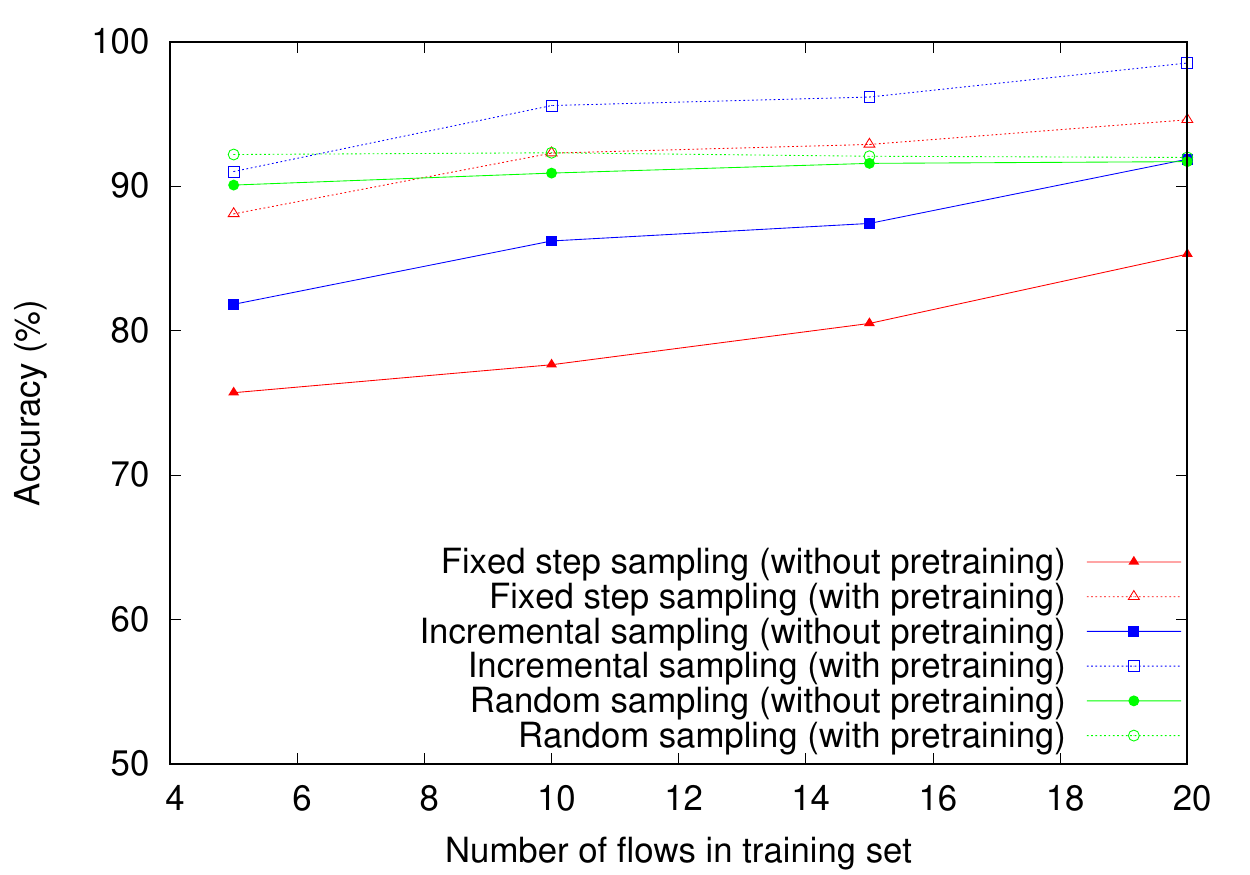}}
\subfigure[]{\includegraphics[width=2.6in]{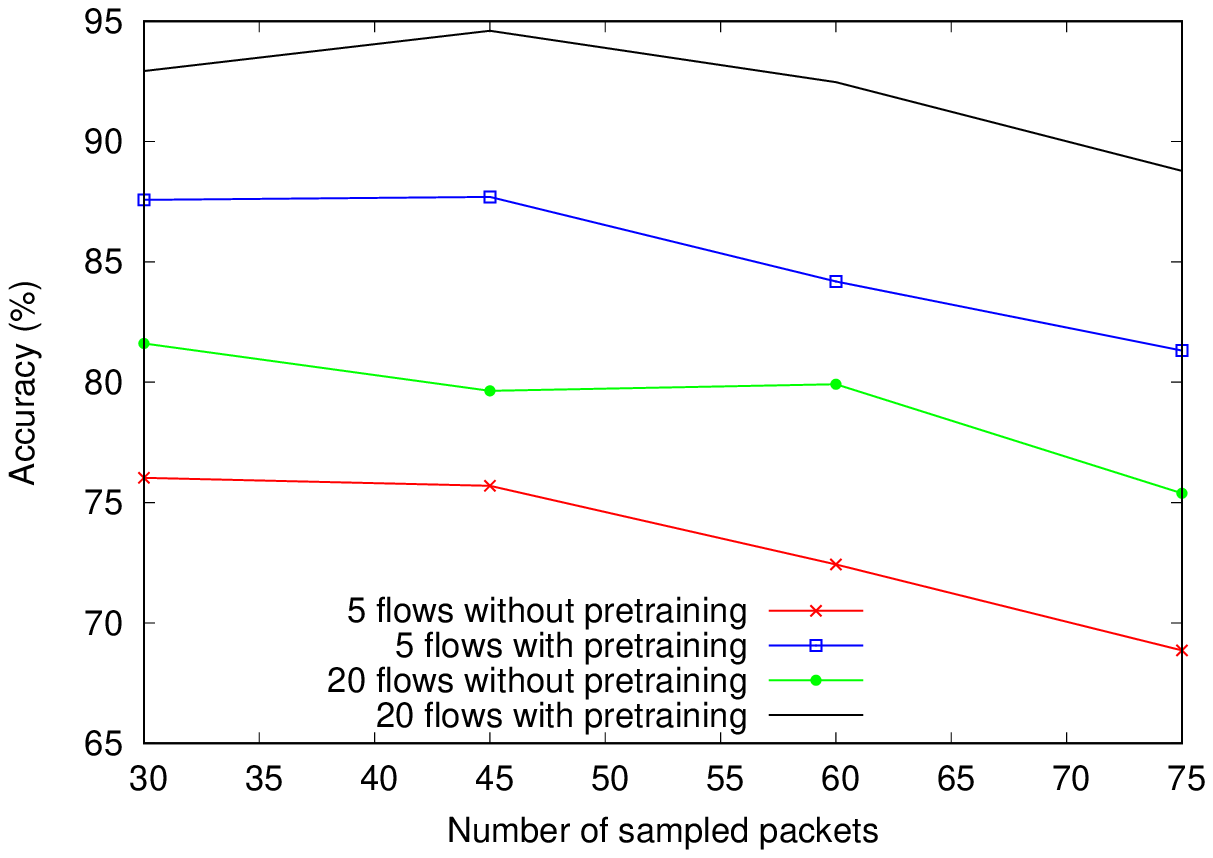}}
\caption{Accuracy of QUIC dataset vs: (a) Supervised training set size (b) number of samples in fixed step sampling}
\label{training-size-vs-accuracy}
\end{figure}

Fig. \ref{training-size-vs-accuracy}(a) presents the accuracy of our model with various settings. It shows that increasing the size of the training set improves the accuracy except for random sampling. The reason is that during random sampling, we always started sampling from the beginning of the flow each $100$ times we sampled a flow. Hence, the data augmentation method with only 5 files gives the model enough data to capture the patterns corresponding to the beginning of the flow. However, the accuracy of random sampling barely increases as the training size increases. We conjecture that this is because imposing randomness during random sampling makes it more difficult for the model to fit to the true distribution. Fixed step and incremental sampling methods work with sampled flows captured from different parts of a flow. Therefore, if different parts of a flow show different patterns, data augmentation of these two methods reveals more patterns. Hence, increasing the training size boosts the accuracy.

As shown in Fig. \ref{training-size-vs-accuracy}(a), incremental sampling outperforms other sampling methods. When random or fixed sampling is used, it is not easy to capture both long and short patterns. Incremental sampling allows sampled flows to contain many packets in short range and some packets in long range. That is why incremental sampling outperforms other sampling methods. Furthermore, the figure clearly shows the efficacy of our transfer learning model on fixed step and incremental sampling, as expected. Our method improves the accuracy around $10\%$ when compared with a model without the pre-training step.

Table \ref{tbl-quic-upperbound-accuracy} represents the accuracy of the CNN model when the entire labeled dataset is used in a supervised manner. In that case, we do not conduct the first step pre-training because the entire dataset is used to train the second model. This gives us the upper-bound for the accuracy of a fully-supervised learning when sampling is used. The accuracy of incremental sampling is close to the upper-bound when only 20 flows per class are used for re-training. But, fixed sampling and random sampling require larger labeled training set during the re-training. To find how much sampling degrades accuracy, we also train a Random Forest (RF) classifier that takes statistical features as input and predicts the class labels. The accuracy of RF is $99.87\%$ which shows that sampling degrades performance up to around $4\%$ for our dataset. Note that we deliberately avoid using deep models such as CNN for this part because the input is a set of statistical features which is not suitable for a shift-invariant model, such as CNN. In our experiment, fully connected neural network is also unnecessarily complex to be trained with our relatively small dataset. Note that our dataset has only around 6500 flows. When using statistical features as an input, it is not possible to augment the dataset. As a result, total number of data points used during training RF was around 100 times fewer than training a model with sampled data.

\begin{table}[t]
	\centering
	\caption{Accuracy of QUIC dataset with different sampling methods 
	}
	\label{tbl-quic-upperbound-accuracy}
	\begin{tabular}{*{15}{c}}
		\hline
		Supervised training size & Fixed step & Random & Incremental \\
		\hline
		20 flows per class & 94.60\% & 91.35\% & 98.53\%  \\
		\hline
		Entire dataset & 96.50\% & 96.92\% & 98.99\%  \\
		\hline
	\end{tabular}
\end{table}

In the second experiment, we study the effect of number of sampled packets on the accuracy of the model. We change the number of sampled packets from 30 to 75. We set the parameters of fixed step sampling method so that it always covers around the range of 1000 packets.
Interestingly, the accuracy drops when we increase the number of sampled packets, as shown in Fig. \ref{training-size-vs-accuracy}(b). Increasing the number of sampled packets improves the accuracy of statistical feature prediction. However, it is harder for the model to learn class labels when input dimension is larger because of the small training set.



\begin{figure}
	\centering
	\includegraphics[width=5.0in]{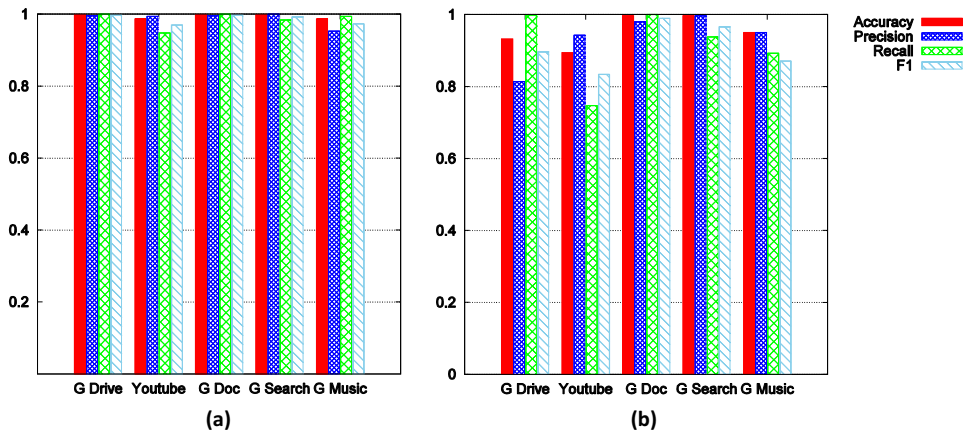}
	\caption{Per class performance metrics of the model on the test set: (a) generated by the scripts (b) generated by human interactions}
	\label{TestHvsS}
\end{figure}

Fig. \ref{TestHvsS}(a) presents the performance metric of our best model, that is, a model re-trained on a pre-trained model using 20 training flows with incremental sampling. The high performance metric shows that it is possible to train a good classifier with as small as 20 flows for each class if our semi-supervised approach is used. Therefore, it dramatically reduces the data collection and labeling that are the most time-consuming and labor-intensive steps.

To study whether automatically generated data with script represents human interaction, we capture 15 flows for each class from interactions of real humans in those 5 Google services. We only use this dataset to test the same model described above. Fig. \ref{TestHvsS}(b) illustrates the performance metrics. Interestingly, accuracy of the Google search and Google document have not changed significantly. However, the accuracy of Google drive, Youtube, and Google music drop up to $7\%$. This depends on how much human interactions can change the traffic pattern, which is class-dependent. Moreover, there are some actions, such as renaming a file or moving files in Google drive, that our scripts do not perform. So, these patterns are not available during re-training. This shows the limitations of datasets and studies \cite{tong2018novel,dubin2017know,conti2016analyzing} that only use scripts to capture data.


QUIC protocol has been introduced in 2012 and Chrome browser has had an experimental support for QUIC since 2014. Therefore, Waikato dataset captured before 2014 cannot have QUIC protocol traffic. Our intuition is that a model pre-trained on Waikato dataset can still be useful for the re-training step because it basically learns how to predict statistical features. It can be considered as a naive statistical feature predictor based on sampled data. Table \ref{tbl-quic-accuracy} presents the accuracy of our method when Waikato dataset is used for pre-training.  Note that sampling method's parameters are different for this experiment. Waikato dataset has many small flows that are not suitable for our previous sampling parameters that covers around 1000 packets. Hence, we change the parameters of sampling methods, similar to the parameter used in next section. That is the reason why the accuracy of even model without pre-training is lower than the experiment shown in Fig. \ref{training-size-vs-accuracy}. Table \ref{tbl-quic-accuracy} clearly shows that the pre-trained model can boost the accuracy even when target traffic does not exist in the pre-training stage. But, the improvement is limited to less than $10\%$ in this case.

\begin{table}[t]
	\centering
	\caption{Accuracy of QUIC dataset with different sampling methods }
	\label{tbl-quic-accuracy}
	\begin{tabular}{*{15}{c}}
		\hline
		Pre-trained on Waikato & Fixed step & Random & Incremental \\
		\hline
		No & 74.51\% & 72.14\% & 74.35\%  \\
		\hline
		Yes & 81.50\% & 81.27\% & 80.76\%  \\
		\hline
	\end{tabular}
\end{table}

\subsection{Ariel Dataset}

We conduct two additional experiments to evaluate the performance of our semi-supervised learning method on public datasets. In the first experiment, we pre-train a model with the unlabeled Waikato dataset to predict statistical features based on sampled flows. Then, we re-train the model with 5 flows of each class in Ariel dataset. We randomly select 5 flows and use the remainder as a test set and repeat the procedure 10 times\footnote{ The reason we do not perform cross-validation is that if we limit each training folds to only contain 5 flows, the total number of cross validation rounds would be too large to be practically evaluated. }. We only show the performance of OS class labels here due to the lack of space. In the second experiment, we pre-train the model with both Waikato and Ariel datasets. Then, we re-train the model similar to the first experiment. When we combine two datasets for pre-training, Ariel flows constitute only around 6\% of the entire combined dataset. We deliberately allow the combined dataset to remain imbalanced to mimic real scenarios where only small portion of unlabeled dataset contains the target labels. Additionally, the parameters we use for fixed sampling, incremental sampling and random sampling are $10$, $(8, 1.2, 10)$, and $0.15$, respectively.

The accuracy of both experiments as well as the one without semi-supervised learning is shown in Table \ref{tbl-Ariel-accuracy}. For brevity, we represent Waikato and Ariel datasets with W and A, respectively. When there is no pre-training, random sampling shows the best accuracy. The similar trend is observed with our QUIC dataset as well, in previous section. Interestingly, there is a small but meaningful gap between the two experiments. That shows even if the target classes are only a small portion of dataset during the pre-training step, they can improve the accuracy. This is useful because the percentage of target task's flows might be probably small in real word when data is captured from an operational network.

\begin{table}[t]
	\centering
	\caption{Accuracy of Ariel datasets with different  configurations}
	\label{tbl-Ariel-accuracy}
	\begin{tabular}{*{15}{c}}
		\hline
		Pre-trained on \textbackslash Sampling & Fixed step & Random & Incremental \\
		\hline
		- & 53.37\% & 70.76\% & 40.37\%  \\
		\hline
		W & 79.76\% & 75.65\% & 80.82\%  \\
		\hline
		W+A & 81.66\% & 78.54\% & 84.53\%  \\
		\hline
	\end{tabular}
\end{table}

\begin{figure}
	\centering
	\includegraphics[width=2.8in]{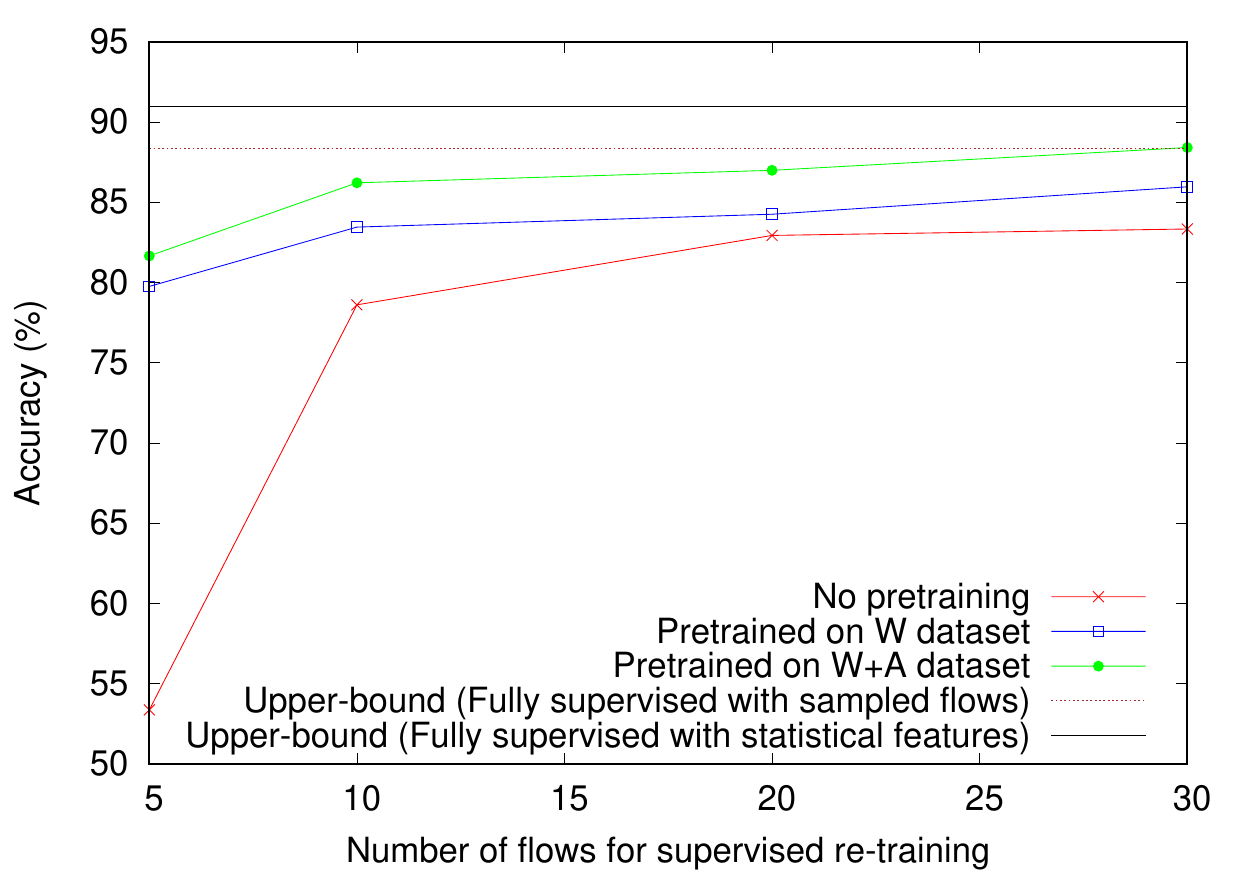}
	\caption{Effect of labeled dataset size on accuracy}
	\label{flows-vs-accuracy-A-dataset}
\end{figure}

To measure the performance gap between semi-supervised and fully-supervised learning, we also train the same CNN-based architecture with the entire A dataset in a fully-supervised manner. First, we train the CNN model with augmented sampled flows from A dataset using fixed step sampling. As it shown in Fig. \ref{flows-vs-accuracy-A-dataset}, the accuracy is around 89\% which can be considered as an upper-bound when sampling is used. To compare how statistical feature prediction degrades the accuracy in comparison with using true statistical features, we perform the following experiment: We feed the true statistical features to the last three layers of the model directly and remove all previous layers. However, the training phase is extremely unstable and high variance with low accuracy. The main reason is that several dense layers of fully connected neural network is too complex to be trained with a small dataset. Recall that when we do sampling, we can sample a single flow multiple times which increases the dataset size. However, in the case of feeding the true statistical features, there is only one set of statistical features for each flow leading to small size dataset. Therefore, we use K-Nearest Neighbor (KNN) for a fully-supervised training with statistical features \footnote{It has shown that it is possible to get a better accuracy with Ariel dataset using some other statistical features \cite{muehlstein2017analyzing}. But, we use the same statistical features that we used as regression targets during the pre-training step to have a fair comparison. }.

The performance gap is shown in Fig. \ref{flows-vs-accuracy-A-dataset}. We only show results of the fixed step sampling due to the lack of space. The trends of other sampling methods are similar. Interestingly, if a pre-trained model contains the target class labels, it can reach the upper-bound accuracy (fully supervised with sampled flows) with only around 30 labeled flows for each class. This is dramatically smaller than typical datasets used for fully-supervised methods in literature. Moreover, even if the unlabeled dataset does not contain target task's flows, the pre-trained model can act as a general function approximation of statistical features because it has already observed a large number of samples during the pre-training step.

\section{Discussion}

Typically, network traffic classifiers use one or combination of the following features \cite{rezaei2018deep}: statistical features, time series, and header/payload contents. The choice of input features depends on many factors, which are comprehensively explained in \cite{rezaei2018deep}. During the pre-training step, our approach takes sampled time series features and outputs statistical features. Hence, our approach does not work on datasets for which time series or statistical features are not good enough for classification. For instance, we also conducted an experiment on ISCX dataset \cite{draper2016characterization} for which the accuracy of model based on statistical and time series features were reported to be around $80\%$  \cite{draper2016characterization}. Our approach failed to produce a model with higher than $68\%$ accuracy when only 20 flows were used from each class during the supervised re-training step. However, a CNN model using payload information achieved above $95\%$ accuracy with fully-supervised learning in \cite{lotfollahi2017deep}.

During the pre-training step, the model see most possible traffic patterns, even the patterns that are not similar to any of the target classes. However, it is possible that during the supervised re-training step, some distinctive patterns are missed from labeled dataset which degrades the accuracy dramatically if the model is used in real environment. Hence, the small labeled dataset should be captured carefully. For instance, it has been shown that user actions in certain Android applications, such as Facebook or Twitter can be identified using time series features \cite{conti2016analyzing}. These actions are sending message, posting status, etc. This means that if target classes are Android applications, one should ensure that all actions are included in the labeled dataset at least once because the pattern of each action is different from another. This is similar to the experiment we did to test our model on human-triggered data where we realized some actions in some Google services did not exist in our training set, such as renaming a file. 

We show that incremental sampling outperforms other sampling methods. Note that we need to choose the parameters appropriately to accommodate the CNN model. CNN uses a set of kernels to cover the entire input. In incremental sampling, the sampling parameter $l$ is increased by $\alpha$ after sampling each $\beta$ packets. If one chooses a large value for $\alpha$, distance between the first few sampled packets are significantly smaller that the last few packets. In that case, CNN model is not suitable because the same filter which is supposed to capture certain pattern for close sampled packets will be used on the far apart samples packets. Hence, when using incremental sampling, one should not use large $\alpha$. 

\section{Conclusion}

In this paper, we propose a semi-supervised learning method that reduces the number of labeled data significantly for network traffic classification. We use 1-D CNN model that takes sampled time series features as input. In the pre-training step, the model is trained to predict statistical features of the entire flow, which does not require human effort for labeling. Then, we transfer the learned parameters to a new model and re-train it with a small labeled dataset. We capture 5 Google services that use QUIC protocol to evaluate our model. We show that with the proposed semi-supervised approach and 20 labeled data from each class the model achieves high accuracy close to a model trained in fully-supervised fashions. We evaluate 3 sampling methods: fixed step, random, and incremental sampling. We also conduct experiments on public datasets to show the generalizability of the proposed approach. We show a model pre-trained on a unlabeled public dataset can improve the accuracy of another labeled dataset. This shows that a model pre-trained with our approach on a large unlabeled dataset can be used as a general traffic classifier that can improve the accuracy of probably any traffic classification tasks if weights are transferred. 

%
%
%
\bibliographystyle{splncs04}

\end{document}